\documentclass{SciPost}

\hypersetup{
    colorlinks,
    linkcolor={red!50!black},
    citecolor={blue!50!black},
    urlcolor={blue!80!black}
}

\usepackage[bitstream-charter]{mathdesign}
\DeclareSymbolFont{usualmathcal}{OMS}{cmsy}{m}{n}
\DeclareSymbolFontAlphabet{\mathcal}{usualmathcal}

\fancypagestyle{SPstyle}{
\fancyhf{}
\lhead{\colorbox{scipostblue}{\bf \color{white} ~SciPost Physics Proceedings }}
\rhead{{\bf \color{scipostdeepblue} ~Submission }}

\fancyfoot[C]{\textbf{\thepage}}
}

\newcommand{\lp}{\left(}
\newcommand{\rp}{\right)}
\newcommand{\lc}{\left[}
\newcommand{\rc}{\right]}

\begin{document}

\pagestyle{SPstyle}

\begin{center}{\Large \textbf{\color{scipostdeepblue}{
Constraining the SMEFT at Present and Future Colliders\\
}}}\end{center}

\begin{center}\textbf{
Eugenia Celada\textsuperscript{1$\star$}
}\end{center}

\begin{center}
{\bf 1} Department of Physics and Astronomy, University of Manchester, Oxford Road, Manchester M13 9PL, United Kingdom
\\[\baselineskip]
$\star$ \href{mailto:eugenia.celada@postgrad.manchester.ac.uk}{\small eugenia.celada@postgrad.manchester.ac.uk}
\end{center}

\definecolor{palegray}{gray}{0.95}
\begin{center}
\colorbox{palegray}{
  \begin{tabular}{rr}
  \begin{minipage}{0.36\textwidth}
    \includegraphics[width=60mm,height=1.5cm]{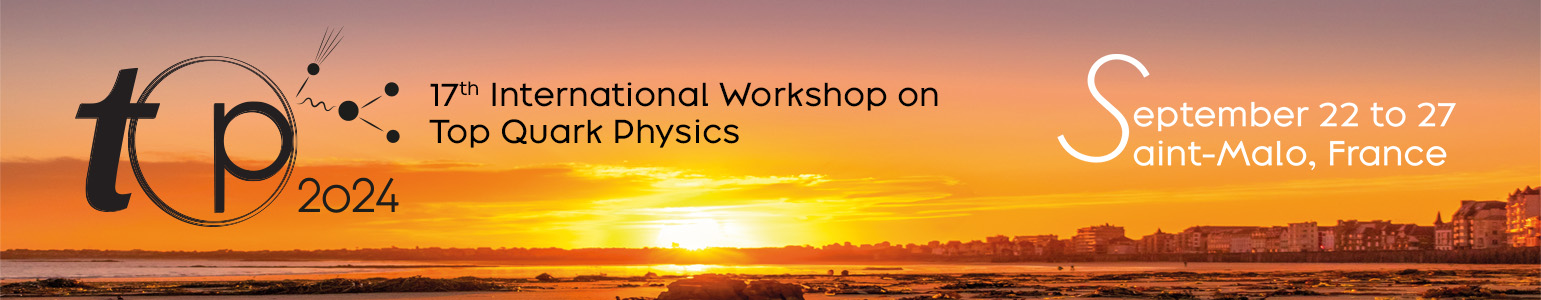}
  \end{minipage}
  &
  \begin{minipage}{0.55\textwidth}
    \begin{center} \hspace{5pt}
    {\it The 17th International Workshop on\\ Top Quark Physics (TOP2024)} \\
    {\it Saint-Malo, France, 22-27 September 2024
    }
    \doi{10.21468/SciPostPhysProc.?}\\
    \end{center}
  \end{minipage}
\end{tabular}
}
\end{center}

\section*{\color{scipostdeepblue}{Abstract}}
\textbf{\boldmath{
We present results from {\sc\small \textbf{SMEFiT3.0}}, a global SMEFT fit of Higgs, top quark, and diboson production data from the LHC. 
Our updated analysis includes recent inclusive and differential measurements from the LHC Run II, together with the exact implementation of electroweak precision observables (EWPOs) from LEP and SLD.
We then analyse the impact of HL-LHC measurements by adding to {\sc\small \textbf{SMEFiT3.0}} the projections obtained by extrapolating from Run II data.
Finally we estimate the potential of two proposed high-energy circular $e^+e^-$ colliders, the FCC-ee and the CEPC, in further improving the bounds on the SMEFT parameters.
}}

\vspace{\baselineskip}

\noindent\textcolor{white!90!black}{%
\fbox{\parbox{0.975\linewidth}{%
\textcolor{white!40!black}{\begin{tabular}{lr}%
  \begin{minipage}{0.6\textwidth}%
    {\small Copyright attribution to authors. \newline
    This work is a submission to SciPost Phys. Proc. \newline
    License information to appear upon publication. \newline
    Publication information to appear upon publication.}
  \end{minipage} & \begin{minipage}{0.4\textwidth}
    {\small Received Date \newline Accepted Date \newline Published Date}%
  \end{minipage}
\end{tabular}}
}}
}

\section{Introduction}
\label{sec:intro}

The end of Run II of the LHC has provided a lot of new measurements, and the High-Luminosity upgrade HL-LHC~\cite{Cepeda:2019klc,Azzi:2019yne}, scheduled after the end of Run III, is expected to accumulate a total integrated luminosity of up to 3 ab$^{-1}$ per experiment.
After the end of its program, several proposals for future particle colliders have been made, including circular electron-positron colliders, such as the FCC-ee~\cite{FCC:2018byv,FCC:2018evy} and the CEPC~\cite{CEPCPhysicsStudyGroup:2022uwl}. 
The interpretation of such a vast amount of data requires a global analysis in order to test the validity of the Standard Model (SM) at the TeV scale and to understand how it can be extended.
The SMEFT is the ideal tool for this purpose, by parametrising high-energy physics effects through precise measurements at low energy that can be revealed in global fits.

In {\sc\small SMEFiT3.0}~\cite{Celada:2024mcf} we performed, within the {\sc\small SMEFiT} framework~\cite{Hartland:2019bjb,Ethier:2021ydt,Ethier:2021bye,vanBeek:2019evb,Giani:2023gfq,terHoeve:2023pvs}, an updated global SMEFT fit at linear and quadratic order in the EFT. 
We considered 50 dimension 6 operators in the Warsaw basis, that can be divided in five classes: two-fermion (2F), the four-fermion classes four-heavy (4H) and two-light-two-heavy (2L2H) and purely bosonic (B), as well as the four-lepton operator $\mathcal{O}_{\ell \ell}$.
Regarding the experimental input, we updated the measurements already included in~\cite{Ethier:2021bye,Giani:2023gfq} with Simplified Template Cross Section (STXS) measurements, the $p_T^Z$ differential distribution in $WZ$, as well as recent top production measurements from ATLAS and CMS. We also upgraded the fit with an exact treatment of the EWPOs, which were approximated in the previous version. 
More details on the theoretical framework and experimental input are given in~\cite{Celada:2024mcf}.
We then considered projections for the HL-LHC, and the two proposed circular colliders FCC-ee and CEPC, and determined their constraining power on the SMEFT coefficients once added on top of the baseline {\sc\small SMEFiT3.0} fit. Having found very similar results for the FCC-ee and CEPC, in this contribution we discuss only the FCC-ee fit results as a representative example.

This work is organised as follows. We discuss in section \ref{sec:2} the updated constraints on the SMEFT parameter space from the global fit to Run II data, and in section \ref{sec:3} the projected sensitivity at the HL-LHC and at the FCC-ee. Finally, in Sect.~\ref{sec:Conclusion} we summarise and discuss future developments.

\section{The {\sc\small \textbf{SMEFiT3.0}} global fit}
\label{sec:2}

\begin{figure}[t]
    \center
    \includegraphics[width=0.445\textwidth]{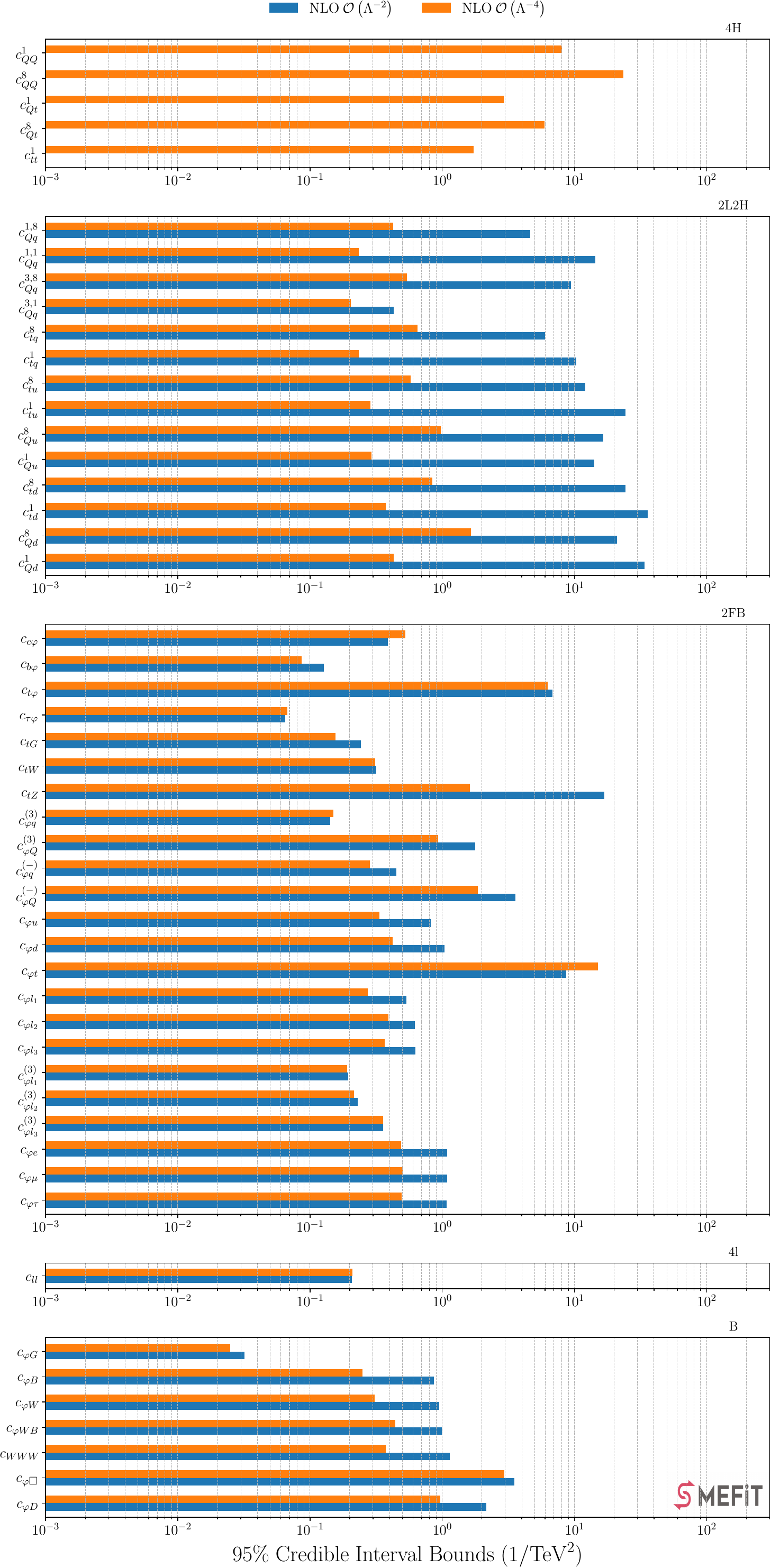}
    \includegraphics[width=0.49\textwidth]{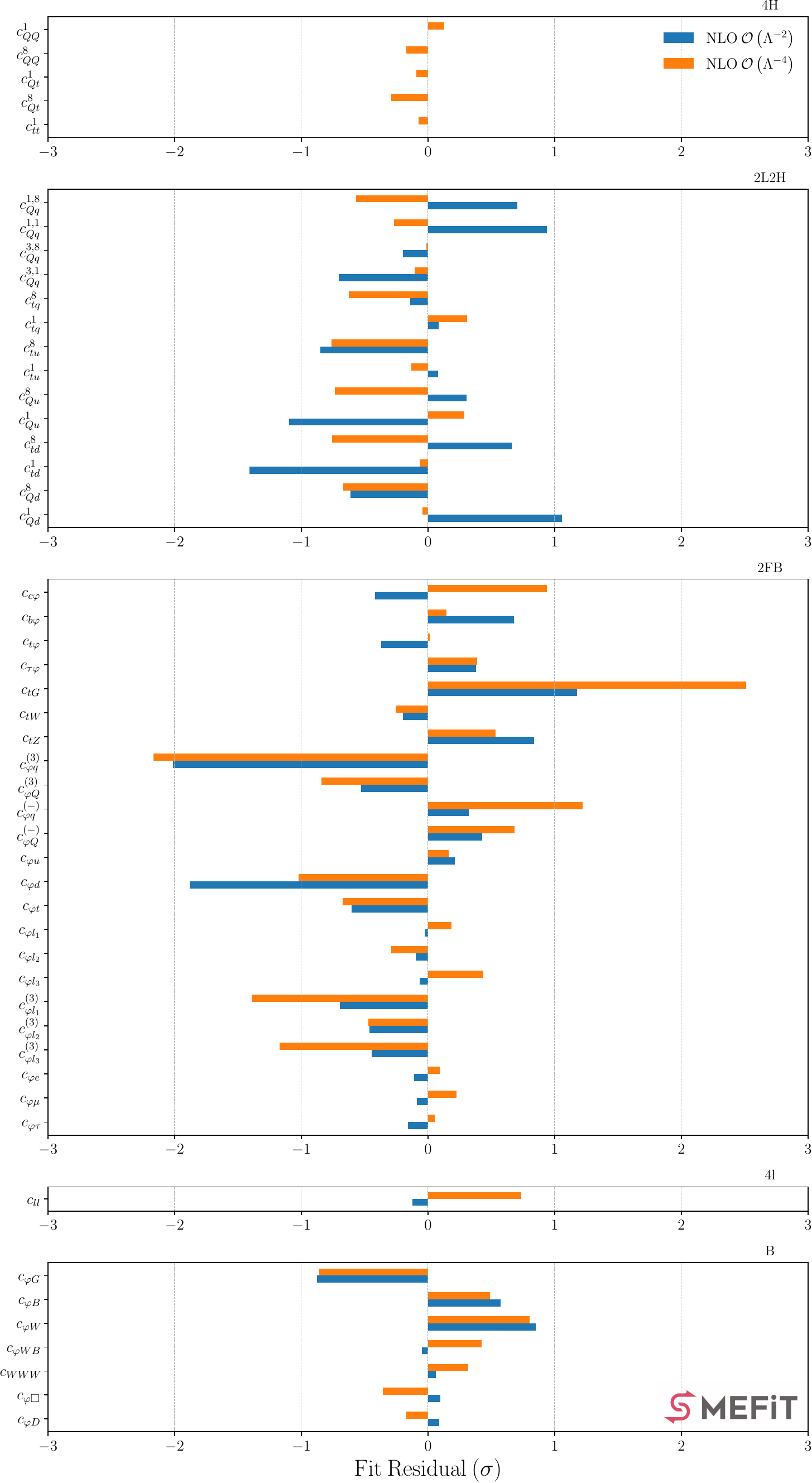}
    \caption{Left: The length of the 95\% credible intervals, expressed in units of 1/TeV$^{2}$, for the $n_{\rm eft}=50$ coefficients entering the fit, both for linear and for quadratic (marginalised) analyses. 
    Right: for the same fits, the residuals between the fit results and the SM expectations defined as in Eq.~(\ref{eq:fit_residuals}). Figure adopted from \cite{Celada:2024mcf}.
    }
\label{fig:smefit30_marginalised_bounds}
\end{figure}

The marginalised bounds on SMEFT coefficients from the current LHC data are shown in the left plot of Fig.~\ref{fig:smefit30_marginalised_bounds}.
For each coefficient, the bar in blue (orange) is the length of the 95\% credible intervals, expressed in units of 1/TeV$^{2}$, in a linear (quadratic) marginalised fit to the updated LHC run II dataset.
The plot shows that most bounds are below 1 and that quadratic terms are in general important and  improve the bounds, in particular for 2L2H operators. They are also necessary to constrain 4H operators, where they break degeneracies by lifting the flat directions present in the linear marginalised fit. However the 4H operators remain the worse constrained with bounds ranging from order 1 to 10.

The agreement with the SM expectations can be quantified with the fit residuals, defined as
\begin{equation}
    \label{eq:fit_residuals}
P_i \equiv 2 \lp \frac{  \left\langle c_i\right\rangle  - c_i^{(\rm SM)}}{
 \lc c_i^{\rm min}, c_i^{\rm max} \rc^{68\%~{\rm CI}}
}\rp \, , \qquad i=1,\ldots, n_{\rm eft}
\end{equation}
so that a residual above 2 indicates a discrepancy with the SM at the 95\% CI. 
These are shown in the right panel of Fig.~\ref{fig:smefit30_marginalised_bounds} and good agreement is found with the SM expectation, with the exception of $c_{tG}$ (only in the quadratic fit), for $c_{\varphi q}^{(3)}$, and for $c_{\varphi d}$ (only in the linear fit). For $c_{tG}$, the large pull arises from a tension between the CMS top-quark double-differential distributions in $(y_{t\bar{t}},m_{t\bar{t}})$ at $8$~TeV and the other ${t\bar{t}}$ measurements included in the fit. The large deviation on the other two coefficients is a consequence of the correlations with other fit parameters.
Finally, we investigated the impact of the new datasets and found that the strongest improvement is on the 2L2H operators, where bounds are reduced by a factor of 2 to 3 and they are driven by the recent $\bar t t$ production measurements.

\section{Prospects at future colliders}
\label{sec:3}

\begin{figure}[t]
    \centering
    \includegraphics[width=0.48\linewidth]{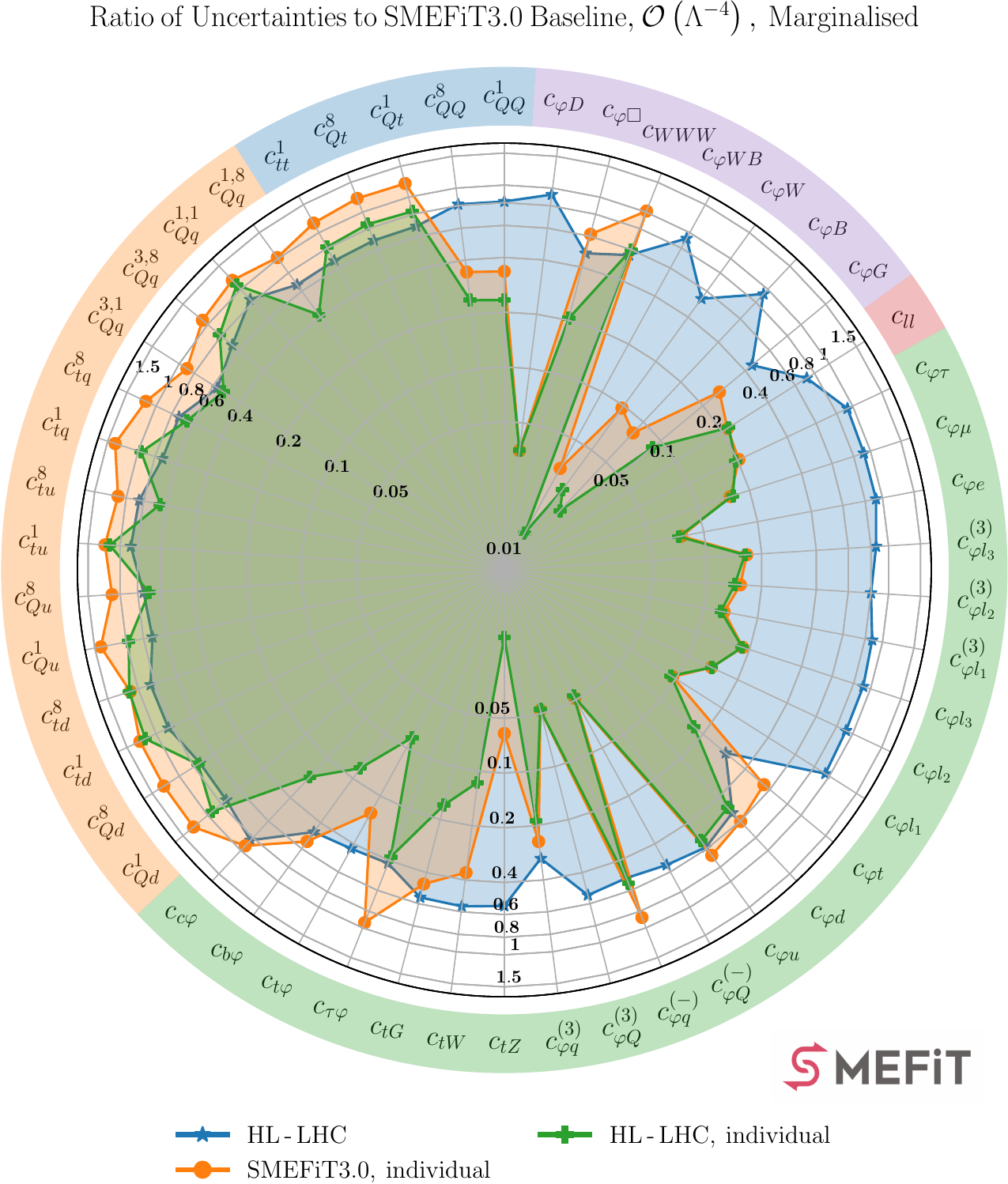}
    \includegraphics[width=0.498\linewidth]{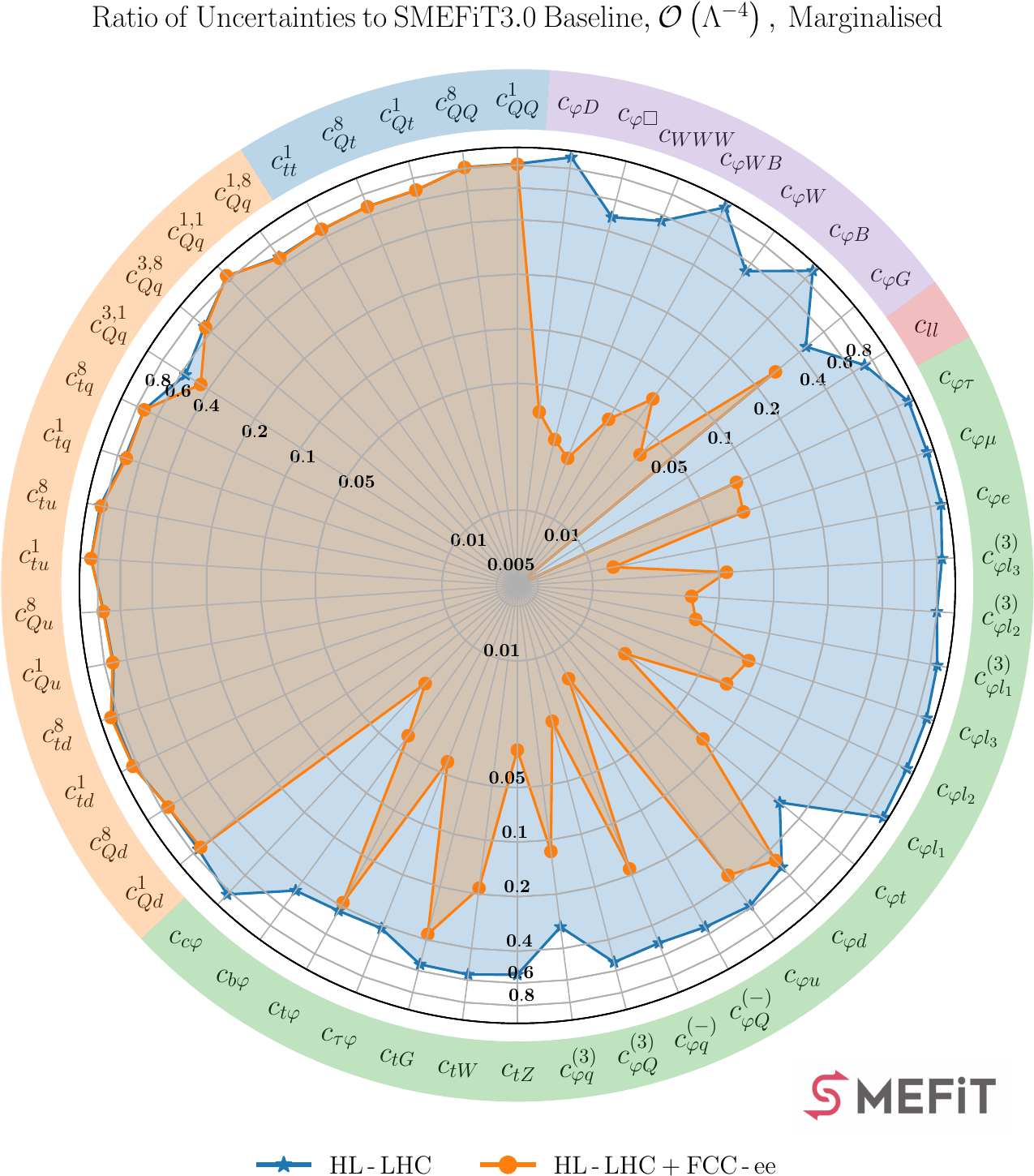}
    \caption{Left: The ratio of uncertainties $R_{\delta c_i}$, defined in Eq.~(\ref{eq:RatioC}), for the $n_{\rm eft}=50$ coefficients entering the quadratic EFT fit, quantifying the impact of the HL-LHC projections when added on top of the {\sc\small SMEFiT3.0}
    baseline. 
    We display both the results of one-parameter fits and those of the marginalised analysis. Right: The ratio $R_{\delta c_i}$ for the marginalised quadratic fits which include first the HL-LHC projections and subsequently both the HL-LHC and the FCC-ee observables. Figure adopted from \cite{Celada:2024mcf}.}
    \label{fig:spider_fcc_nlo_quad_glob}
\end{figure}

We then used {\sc\small SMEFiT3.0}  as a baseline fit to assess the impact of the HL-LHC and FCC-ee projections.
The fit results for future colliders are presented in the two plots of Fig. \ref{fig:spider_fcc_nlo_quad_glob}, where we display the ratio between the magnitude of the 95\% CI for a given EFT coefficient $c_i$, to that of the same quantity in the baseline fit:
\begin{equation}
    \label{eq:RatioC}
R_{\delta c_i} = \frac{ \lc c_i^{\rm min}, c_i^{\rm max} \rc^{95\%~{\rm CI}}~({\rm baseline+\text{HL-LHC}})}{
 \lc c_i^{\rm min}, c_i^{\rm max} \rc^{95\%~{\rm CI}}~({\rm baseline})
} \, , \qquad i=1,\ldots, n_{\rm eft} \, ,
\end{equation}
such that for each $c_i$ the smaller $R_{\delta c_i}$ the higher the improvement with respect to the baseline.
The projections for HL-LHC measurements are obtained by fluctuating the central value around the SM within the experimental uncertainty, while the projected statistical and systematic uncertainties are obtained rescaling the LHC Run II ones by the luminosity and  by a factor of 2 respectively.
The improvement on quadratic marginalised bounds after adding HL-LHC projections, in blue on the left of Fig. \ref{fig:spider_fcc_nlo_quad_glob}, ranges between 20\% to a factor of 3. 
The comparison between the quadratic marginalised and individual bounds displayed by the orange and green lines indicates that in the individual fits the sensitivity is typically much better than in the global fit, in some cases by more than an order of magnitude.
Further improvement is also expected thanks to the increased statistics of the HL-LHC measurements, which would allow for a finer binning and an extended range of differential distributions. Such optimisation was not considered in this analysis and is left for future work.

Regarding the FCC-ee projections, four running scenarios were considered, starting from the $Z$-pole all the way up to $\sqrt{s}=365$~GeV, above the top-quark pair production threshold. we included the following observables that become accessible at these energies: the EWPOs at the $Z$-pole; light fermion pair production; Higgs boson production in both the $hZ$ and $h\nu\nu$ channels; gauge boson pair production;  and top quark pair production.
The blue and orange lines in the right plot of Fig.~\ref{fig:spider_fcc_nlo_quad_glob} show the ratio of the bounds for marginalised quadratic fits where HL-LHC and FCC-ee projections are subsequently included. The plot highlights the substantial improvement of the FCC-ee projections, in particular on the bosonic and two-fermion operators, whose bounds improve by a factor of 30 to 50. In the four-quark sector on the other hand the improvement is limited and only originates as a consequence of marginalisation. Indeed, all the observables considered for the FCC-ee are unsensitive to these operators at LO, and the inclusion of NLO effects is left for future work.

\section{Conclusion}
\label{sec:Conclusion}

In this work we presented {\sc\small SMEFiT3.0}, an updated global fit to LHC data, including the recent Higgs, top and diboson measurements from the LHC Run II. We found that, assuming a high scale of 1/TeV$^{2}$, the current constraints are of $\mathcal{O}(1)$ for most coefficients, and that quadratic SMEFT effects play a fundamental role in reducing the bounds by breaking correlations among them.
This fit was then used as a baseline to assess the impact of future measurements at the HL-LHC and at a future circular electron positron collider, the FCC-ee. 
We found that the marginalised bounds are expected to improve by a 20\% to a factor of 3 after the HL-LHC run.
The inclusion of FCC-ee projections after the HL-LHC would further improve the bounds by up to two orders of magnitudes on coefficients of bosonic and two-fermion operators.
We finally aim to improve the present analysis in several ways. These include improving the HL-LHC projections by considering a dedicated binning. We plan to include projections for other proposed future collider, such as the ILC, CLIC and the high-energy muon collider, and provide a complete comparison of their constraining power.
Finally, the impact of coefficient mixing due to the RGE running is also to be considered.

\paragraph{Funding information}
This work is supported by the European Research Council (ERC) under the European
Union’s Horizon 2020 research and innovation programme (Grant agreement No. 949451).

\bibliography{SciPost_Proceedings_TOP2024_Template.bib}

\end{document}